\documentclass[twocolumn,showpacs,preprintnumbers,amsmath,amssymb]{revtex4-1}

\usepackage{graphicx}
\usepackage{bm}

\begin{document}

\title{ Wetting transition in the McCoy-Wu model}

\author{X. T. Wu}

\affiliation{Department of Physics, Beijing Normal University,
Beijing, 100875, China}

\date{\today}

\begin{abstract}
The wetting transition is studied in the McCoy-Wu Ising model in which the random bonds are perfectly correlated in the direction parallel to the walls . The model is solved numerically on finite size lattices up to $200 \times 200^2$. It is shown that the wetting transition is first-order. For a fixed surface field, the distribution of wetting transition temperature is obtained from $1000$ samples. The results show that the deviation of the wetting transition temperature does not decreases as the lattice size increases. It is shown that for a fixed surface field the wetting transition temperature is sample dependent even in the thermodynamic limit.
\end{abstract}


\maketitle

\section{INTRODUCTION}
The disorder is ubiquitous in nature. The influence of quenched disorder on the phase transition has been of great interest in the theoretical and experimental physics \cite{vojta}.  In the three dimensional Ising model, the quenched disorder leads to a new critical point with exponents different from the pure ones  \cite{landau,diluteising}. In the McCoy-Wu model, the disorder causes the Griffiths-McCoy singularity, where some thermodynamic quantities are singular in a range of temperature, rather than just at the critical point   \cite{mccoy,mccoy1,fisher}. It is shown rigorously that in 2D quenched randomness results in the suppression of first-order phase transitions in the random-field Ising model, random bond Potts model and spin glasses \cite{aizenman}. 

In the present work we show two novel effects of the disorder on the phase transition. We study the wetting transition in the McCoy-Wu Ising model with surface fields, which is depicted in Fig. 1. In this model on the two dimensional lattice, all the vertical bonds are the same, while the horizontal bonds are identical to each other within each column but differ from column to column \cite{mccoy}.   The first effect is that the phase transition is  first-order even with the disorder. The second one is that for a fixed surface field the wetting transition temperatures are sample dependent and do not to converge to a limit as the size of the system goes to infinity.  To our knowledge, in the all previous studied disordered systems the phase transition temperatures converge to a limit as the system size goes to infinity (see  examples in Ref. \cite{domany,chakravarty,bellafard}). This should be the strongest effect of the disorder on the phase transition discovered up to now.   

Wetting transition on the Ising model was first studied by Abraham with the two-dimensional Ising model, in which two opposite external field are applied on  the boundary \cite{abraham}. In the  Abraham's exact solution, it is shown that the wetting transition is continuous and the average distance of the interface (separating the predominantly $+$ and $-$phases) from the boundary diverges smoothly to infinity. Forgacs, Svrakić and Privman found that the wetting transition becomes first-order if one adds a line defect in the bulk \cite{privman}. The McCoy-Wu model can be regarded as an Ising model being added many line defects. 

\begin{figure}
\includegraphics[width=0.5\textwidth]{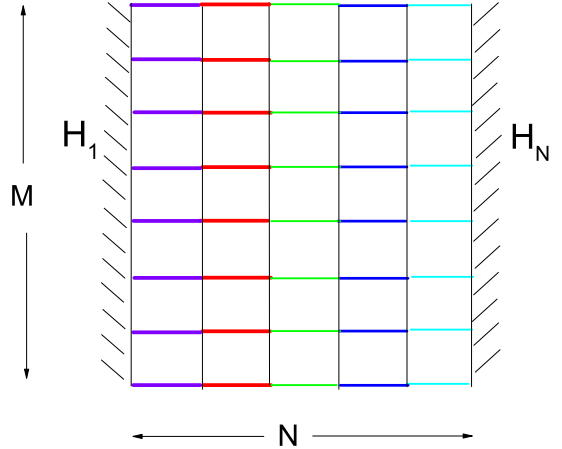}
\caption{ (a) The sketch of the McCoy-Wu model with surface fields. The surface fields act on the spins at the first and last column. The first column is usually called the left wall. The horizontal segments in different colors represent random horizontal bonds. }

\end{figure}

We solve the model on finite size lattices with the Bond Propagation Algorithm (BPA) and Site Propagation Algorithm (SPA) \cite{loh,wu1,wu}. We reveal the physical picture of the wetting transition. It is found that the thermodynamics at the wet phase is dominated by the group with most adjacent line defects and at the nonwet phase by the bonds near the boundary. This leads to the  sample dependent wetting transition temperature and the distribution width of the wetting transition temperature should remain finite even if the size of lattice goes to infinity.

The paper is orgnized as follows. In Sec. II, we define the model and introduce the numerical method, BPA and SPA. In Sec. III, the evidences of first-order transition are shown. In Sec. IV, we obtain the distribution of wetting transition temperature on the finite size lattices. In Sec. V, we reveal the physical picture of the wetting transition and argue that the transition temperature is sample dependent in the thermodynamic limit.  In Sec. VI, we propose two semi-random models to improve our arguments further. Sec. VII is a discussion and acknowledgment.

\section{THE MODEL}

The McCoy-Wu model with surface fields is sketched in Fig. 1, in which  all the vertical bonds are the same, while the horizontal bonds are identical to each other within each column but differ from column to column. Consider a set of spins $\sigma(n,m)=\pm 1$ located at points $(n,m)$ of the planar square lattice such that $1\le n \le N, 1\le  m \le M$. The energy of a configuration $\{ \sigma \}$ of spins is given by
\begin{eqnarray}
E & = & -J\sum_{m=1}^{ M-1}\sum_{n=1}^{N} \sigma_{n,m}\sigma_{n,m+1} \nonumber \\
& &-J\sum_{m=1}^{ M}\sum_{n=1}^{N-1} a_n\sigma_{n,m}\sigma_{n+1,m} \nonumber \\
  & &-\sum_{m=1}^{M}[H_1\sigma_{1,m}+H_N\sigma_{N,m}]
\label{eq:lattice1}
\end{eqnarray}
where $H_1$ and $H_N$ are the surface field. The left and right boundary are often referred to competing walls. The random bond $a_n J$ is the horizontal bond in the $n$th column. In this paper we consider the binary bond disorder probability distribution:
\begin{equation}
p(a_n)=\frac{1}{2}[\delta(a_n-1)+\delta(a_n-0.9)].
\end{equation}
The horizontal  bonds are either strong $J$  or weak $0.9J$  with half probability. We use this special case as an example, but the conclusions should be general.  The normalized canonical probability is $P(\{ \sigma \})=Z^{-1}\exp [-\beta E]$ where $Z$ is the canonical partition function. We set $J/k_B=1$ , where $k_B$ is the Boltzman constant. Following the convention, we call a given configuration of disorder $(\{ a_n \})$ a sample. For a given sample, we calculate this model with BPA and SPA \cite{loh,wu1,wu}. 

We consider two types of boundary conditions as done in Abraham's model \cite{abraham}:
\begin{eqnarray}
+-: & ~~~H_N & =1,~~~H_1=-a_0  \nonumber \\
++: & ~~~H_N & =1,~~~H_1=a_0
\label{eq:bound}
\end{eqnarray}
where $a_0>0$. Throughout this paper,  we set 
\begin{equation}
a_0=0.4
\end{equation}
in the numerical calculation. Under the boundary condition $+-$, there is an interface. Then the interfacial free energy (density) is defined
\begin{equation}
f=-\frac{k_B T}{M}\ln \frac{Z_{+-}}{Z_{++}}
\label{eq:interfacial}
\end{equation}
Fixing surface field $a_0$, there is a wetting transition for the boundary condition $+-$ as the temperature changes. There is no singularity in $\ln Z_{++}$. The subtraction of $\ln Z_{++}$ just eliminates the non-singular background term in $\ln Z_{+-}$.  

Correspondingly, we define the interfacial internal energy (density)  by 
\begin{equation}
 u=\frac{\partial (\beta f)}{\partial \beta} 
\end{equation}
 and the interfacial specific heat  by 
\begin{equation}
 c= \frac{\partial u}{\partial T} . 
\end{equation}
These quantities can be calculated with BPA on finite-size lattices.

The magnetization $\overline{\sigma}_{n,m}$, which is the thermodynamic average of the spin at the site $(n,m)$, is defined by
\begin{equation}
\overline{\sigma}_{n,m}=\sum_{\{\sigma_{l,k}\}}\sigma_{n,m}e^{-E(\{\sigma_{l,k}\})/K_BT}.
\end{equation}
Note that the above average for the magnetization is carried on a given sample, a specific disorder configuration. It is not the average over many samples. 

Note that the top and bottom boundary are open in this model because of our algorithm. In our numerical calculation, we set 
\begin{equation}
M=N^2.
\end{equation}
Moreover it has $N \ge 80$. So the lattices with size $N \times M$ are very narrow rectangles. Hence the effect of the open boundary at the top and bottom can be ignored. 

We solve the model on finite size lattices with the Bond Propagation Algorithm (BPA) and Site Propagation Algorithm (SPA) \cite{loh,wu1,wu}. These algorithms are very accurate and can be carried out on very large lattices. In our numerical calculation, the largest size of the lattice is $200 \times 200^2$. One can calculate the free energy, internal energy and specific heat with BPA, and the magnetization on each site with SPA. These calculation are in a high accuracy with error less than $10^{-6}$.  

\section{The evidence for the first order phase transition}

In this section, we show the wetting transitions in some samples on the lattice with size $N=120$ and $N=200$. It is shown that the interfacial specific heat in these samples satisfy the scaling function of first-order phase transition.

\begin{figure}
\includegraphics[width=0.5\textwidth]{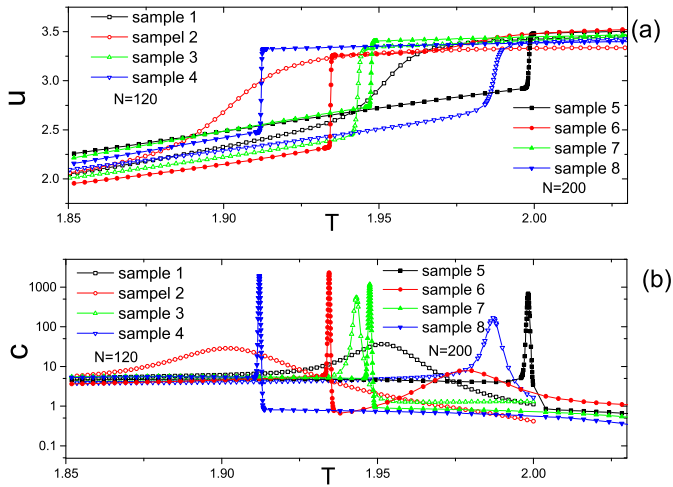}
\caption{ (a) The interfacial internal energy for four samples with $N=120$ and four samples with $N=200$. (b) The corresponding interfacial specific heat for the eight samples}

\end{figure}

Fig. 2(a) shows the interfacial internal energy and specific heat for four samples with $N=120$ (sample $1,2,3,4$) and four samples with $N=200$ (sample $5,6,7,8$). They have a jump in the interfacial interanl energy as temperature changes. The jumping in $u$ is far more drastic for the sample $5,6,7,8$ with $N=200$ than the sample $1,2,3,4$ with $N=120$. Figure 2(b) shows the  corresponding interfacial specific heat for the eight samples.  The interfacial specific heats for samples $5,6,7,8$ with $N=200$ show very narrower and higher peaks than samples $1,2,3,4$ with $N=120$. 

The wetting transition temperatures $T_w$  is defined as at which the interfacial specific heat is maximal. $T_w$ should be actually called the pseudo-phase-transition temperature because the systems in our calculation are finite.  To be simple, we call $T_w$ the wetting transition temperature in the following. The maximum of the interfacial specific heat is denoted by $c_m$. As shown in Fig. 2, the wetting temperatures $T_w$ are different for sample to sample.  Moreover, the difference between the wetting transition temperatures for any two samples is usually much larger than their widths of the specific heat peaks. 

We find that the interfacial specific heat peak satisfies the the following scaling function, which is a typical finite-size scaling for the first order \cite{binder1}:
\begin{equation}
\frac{c}{c_m}=[(e^{t}+e^{-t})/2]^{-2}
\label{eq:gauss}
\end{equation}
where 
\begin{equation}
t=\frac{T-T_w}{\tau}
\end{equation}
and $\tau$ characterizes the width of the interfacial specific heat peak, which is obtained by fitting the numerical results. 

Fig. 3 shows the collapses of  interfacial specific heat for the eight samples rescaled with Eq. (\ref{eq:gauss}). The parameters $c_m, T_w, \tau $ are given in Table I. We can see the main part of the interfacial specific heat peak are perfectly given by the scaling function Eq. (\ref{eq:gauss}), although the parameters of $c_m,\tau$ for the samples differ very much.

\begin{figure}
\includegraphics[width=0.5\textwidth]{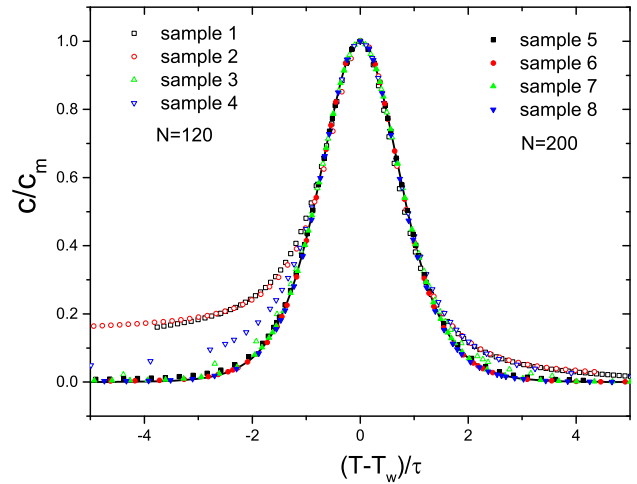}
\caption{ Scaling the interfacial specific heat for the eight samples shown in Fig. 2(b). The full curve represents Eq. (\ref{eq:gauss}).}

\end{figure}

\begin{table}[htbp]
 \caption{ The parameters of $c_m, T_w, \tau $ for the Fig. 2(b). }
\begin{tabular}{cccl}
\hline
      & $c_m$ & $T_w$ & $\tau$  \\
 sample 1~~~~~ & $36.5774$~~ & $1.951219$~~ & $0.0117$ \\
 sample 2~~~~~ & $28.5872$~~ & $1.901140$~~ & $0.0214$ \\
 sample 3~~~~~ & $547.562$~~ & $1.943223$~~ & $0.000791$ \\
 sample 4~~~~~ & $167.495$~~ & $1.987095$~~ & $0.00184$ \\
 sample 5~~~~~ & $702.838$~~ & $1.999803$~~ & $0.000401$ \\
 sample 6~~~~~ & $2291.39$~~ & $1.934441$~~ & $0.000214$ \\
 sample 7~~~~~ & $1166.27$~~ & $1.947504$~~ & $0.000298$ \\
 sample 8~~~~~ & $1915.94$~~ & $1.912100$~~ & $0.000232$ \\

\hline
\end{tabular}
\label{a1}
\end{table}

The agreement with Eq. (\ref{eq:gauss}) are much better for the four samples with $N=200$, sample $5,6,7,8$ than the four samples with $N=120$, sample $1,2,3,4$ in Fig.2. In the samples with $N=200$ the surface specific heat peaks $c_m$ are much higher and the widths $\tau$ are much narrower than those with $N=120$. As shown later, as the system size grows, the average of $c_m$ diverges and the average of $\tau$ converges to zero. This infers that the interfacial specific heat for the larger system can be better described by Eq. (\ref{eq:gauss}), which is a typical finite-size scaling for the first order \cite{binder1}. This is the first clue that this wetting transition is first-order.

Here we give a derivation of the scaling function Eq. (\ref{eq:gauss}). It can be obtained by following Binder's argument on the finite size scaling of the first order transition in Potts model. See the argument leading to the equation ($2.41$) in the reference \cite{binder1}. Letting $q=1$ and $c_{+}/c_{-}=1$ give rise to Eq. (\ref{eq:gauss}). In reference \cite{binder1}, q-state Potts model is discussed and there are q-fold degenerate ordered states. Here both below and above the wetting transition the states are ordered and there is only one state, so one can let $q$ to be $1$. The wetting transition takes place under the boundary condition $+-$, and $c_+$ and $c_-$ should be the specific heat for the whole system below and above the wetting transition respectively. Then the heat capacity difference below and above the transition is given by $(c_+-c_- )MN$.  On the other hand, the heat capacity difference below and above the transition equals to  $M\Delta c$, where $\Delta c$ is the difference of interfacial specific heat  below and above the transition.  Therefore it has $(c_+-c_- )MN=M\Delta c$, so $c_+ -c_- =M\Delta c/(MN)=\Delta c/N$. Obviously the difference of the interfacial specific heat below and above the transition $\Delta c$ should be finite, so in the limit of thermodynamic limit, i.e. $N \rightarrow \infty$, $c_+ -c_-$ vanishes. Therefore in the limit of thermodynamic limit, $c_+/c_-$ equals $1$. 

We calculate the magnetization $\overline{\sigma}_{n,m}$ at the site $(n,m)$ to obtain the magnetization profile.
Because the system is translation invariant in the vertical direction, the magnetization $\overline{\sigma}_{n,m}$  only depends on the x-coordinate "$n$". So, the magnetization in the middle row $\overline{\sigma}_{n,M/2}$, $1\le n \le N$ can represent the magnetization profile as shown in Fig. 4 (a). From the magnetization profile, we can get the position $x_d$ of the interface (or domain wall, where the magnetization is zero. It is obtained by a simple interpolation between the magnetization at the $n_d$th column and $n_d+1$th column, 
\begin{equation}
x_d=n_d+\frac{\overline{\sigma}_{n_d,M/2}}{\overline{\sigma}_{n_d,M/2}-\overline{\sigma}_{n_d+1,M/2}}
\end{equation}
if $\overline{\sigma}_{n_d,M/2}<0$ and $\overline{\sigma}_{n_d+1,M/2}>0$, the magnetization changes sign. We also call $x_d$ the interface position.

\begin{figure}
\includegraphics[width=0.5\textwidth]{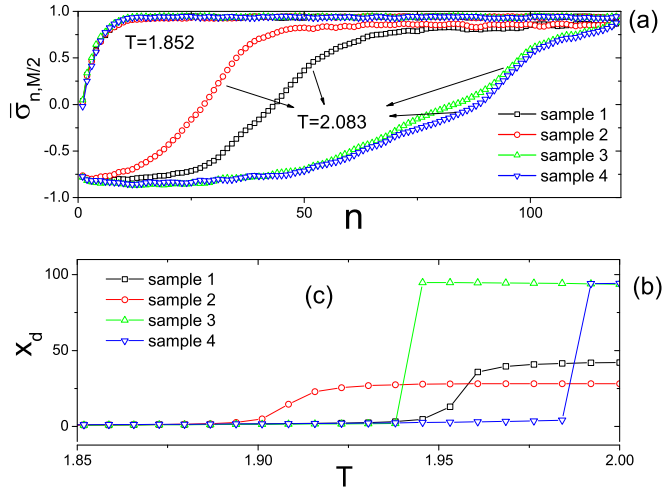}
\caption{ (a) The magnetization profile for the four samples $1,2,3,4$ in Fig. 2,3 at two temperatures. At $T=1.852$ the four samples are at nonwet phase and at $T=2.083$ they are at wetting phase. (b) The interface position $x_d$ vs the temperature.}
\end{figure}

In Fig. 4, we show the magnetization profiles and the interface positons for samples $1,2,3,4$ in Fig. 2 and 3. Fig. 4(a) shows the magnetization profiles of the four samples at two temperatures. At $T=2.083$, the systems are at the wet phase, where the interfaces are far from the left wall. At $T=1.852$, the systems are at the nonwet phase, where the interfaces are pinned at the left wall. Fig. 4(b) shows the interface postion v.s. temperature for the four samples. As the temperature increases the interface unbinds discontinuously from the left wall, where $n=0$, and becomes localized far from the left wall. The distance of the interface from the left wall $x_d$ has a jump at the wetting transition point. The rounding of the jumping of  $x_d$ is due to the finite size effect. The temperatures at the jump of the interface position $x_d$ coincide with those at the maximal interfacial specific heat in Fig. 2. 

\section{Finite size scaling of the average quantities and their deviations}

Since the physical quantities $T_w,c_m,\tau$ depends on the configuration of the disorder, we study their distribution over more than $1000$ disorder configurations for $N=80,120,160,200$. For each configuration of disorder on a finite lattice, we calculate the interfacial specific heat at different temperatures. To search the wetting transition point precisely and efficiently  we adopt the following procedure. At first, we calculate the specific heat at five points with equal distance with the temperature lower bound being $1.78$ and the upper bound being $2.12$. The transition usually takes place in this range. Then compare the specific heats of the middle three points and pick out the one at which the specific heat is maximal. Take this point as the central point and the two neighbored points as the lower and upper bounds and repeat the first step again. After repeating this procedure $13$ times, we take the point at which the specific heat is maximal as the transition point $T_w$ and the specific heat at this point as specific heat maximum $c_m$. Then the precision of $T_w$ is about $10^{-5}$. We use the  data obtained during the iterations to get $c_m$ and $\tau$ with Eq. (\ref{eq:gauss}).

\begin{figure}
\includegraphics[width=0.5\textwidth]{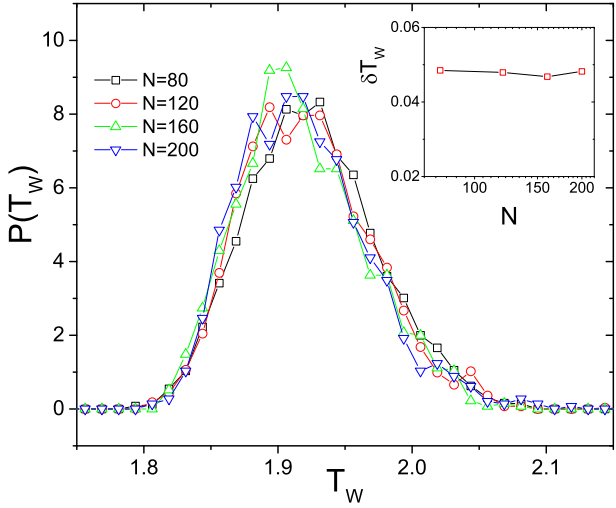}
\caption{ The distribution of wetting transition temperature and its distribution width for $N=80,120,160,200$. The inset shows the deviations of the transition temperatures for $N=80,120,160,200$.}

\end{figure}

Figure 5 shows the distribution of wetting transition temperature $T_w$ for $4$ lattice sizes $N=80,120,160,200$. The number of samples, the disorder configurations, is $N_s=3956,1191,1080,1170$ for $N=80,120,160,200$ respectively. The distribution $P(T_w)$ is defined as $P(T_w)=n_{T}/N_s$ where $n_{T}$ is the number of samples with $T<T_w<T+\Delta T$. 

The average of the wetting transition temperature is obtained by $\overline{T}_w=\sum_i^{N_s}T_{wi}/N_s$, where $T_{wi}$ is the transition temeperature of the $i$th sample. For the system size $N=80,120,160,200$ the averages of transition temperature are $\overline{T}_w=1.93(3),1.92(6),1.92(6),1.92(6)$ respectively. The average of the transition temperature keeps the same as the lattices size increases.

The deviation of the wetting transition temperature is obtained by $\delta T_w=\sqrt{\sum_i^{N_s}(T_{wi}-\overline{T}_w)^2/N_s}$. The inset of Fig. 5 shows the deviations of the wetting transition temperatures for lattice size $N=80,120,160,200$ . One can see that the deviation $\delta T_w$ does not decrease as the lattice size increases.  It is not strange for the averaging transition temperature $\overline{T}_w$ not to change with the lattice size. But it is strange for the deviation also not to change with the lattice size.

\begin{figure}
\includegraphics[width=0.5\textwidth]{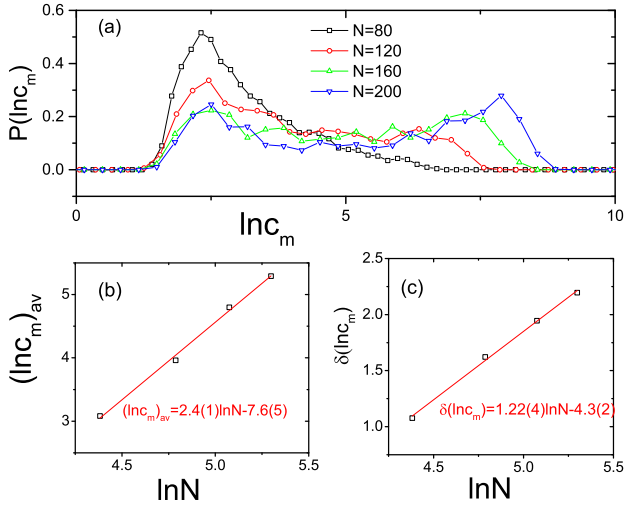}
\caption{ The distributions of interfacial specific heat maximum for $N=80,120,160,200$ are presented in (a).  (b)The average of $\ln c_m$ vs the system size. (c)The deviation of $\ln c_m$ vs the system size.}

\end{figure}

Now we discuss the statistics of $c_m$, the maximal interfacial specific heat. Since $c_m$ are distributed in wide scale, we calculate the distribution of $P(\ln c_m)$, which are shown in Fig. 6. As one can see, the qualitative properties of the distributions are different. For $N=80$ the distribution has a single sharp peak. For $N=120$, the distribution has a long tail. For $N=160,200$, it has two peaks, one at the smaller $c_m$ side and another one at the larger $c_m$ side. Obviously, we can not simply rescale these distributions to collapse. 

Because the distibution of specific heat maximums is broad, we study the distribution of their logarithmic, i.e. $P(\ln c_m)$, which is defined as $P(\ln c_m)\Delta \ln c_m=n_{c}/N_s$ where $n_{c}$ is the number of samples satisfying $\ln c_m<\ln c_{mi}< \ln c_{m} +\Delta \ln c_m$. Fig. 6(a) shows the distribution of $P(\ln c_m)$ for lattice size $N=80,120,120,160,200$. Their averages are defined by $(\ln c_m)_{av}=\sum_i^{N_s}\ln c_{mi} /N_s$ and  deviation are defined by $\delta (\ln c_m)=\sqrt{\sum_i(\ln c_{mi}-(\ln c_m)_{av})^2/N_s}$. They are shown in Fig. 6(b) and 6(c).  $\ln c_{mav}$ and $\delta (\ln c_m)$ vs $\ln N$ lie almost in  straight lines. Simply linear fitting yields 
\begin{equation}
(\ln c_m)_{av}=2.4(1)\ln N-7.6(5).
\end{equation}
and 
\begin{equation}
\delta (\ln c_m)=1.22(4)\ln N-4.3(2).
\end{equation}
The average of $\ln c_m$ and its deviation increase with the lattice size $N$.

\begin{figure}
\includegraphics[width=0.5\textwidth]{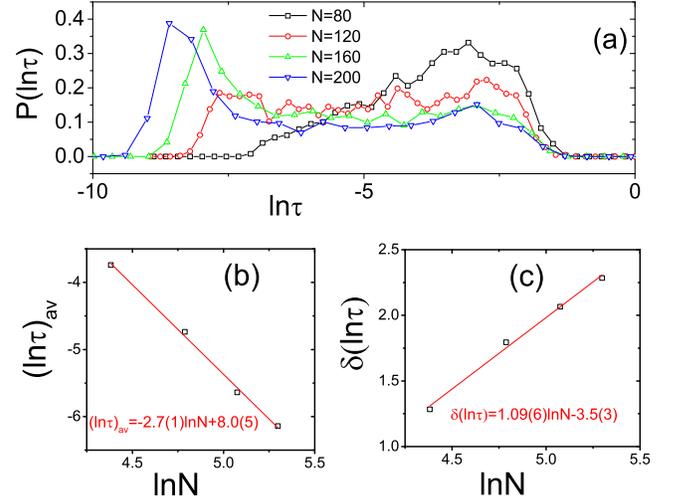}
\caption{ The distributions of specific heat peak width  for $N=40,60,80,100,120,160,200$ are presented in figure (a)-(g). Figure (h) shows the average of the specific heat peak width and it deviation vs the system size}

\end{figure}

For the width of the specific heat peak $\tau$, we calculate the distributions of $P(\ln \tau)$. They are shown in Fig. 7. The distributions seem to belong three types. For $N=40,60,80$ the distribution has a single sharp peak. For $N=100,120$, the distribution has a plateau. For $N=160,200$, it has a plateau and a peak, one at the smaller $\tau$ side.

However their averaged width $(\ln \tau)_{av}=\sum_i^{N_s}\ln \tau_i$ over $N_s$ samples and its deviation $\delta (\ln \tau)=\sqrt{\sum_i(\ln \tau_i-(\ln \tau)_{av}))^2/N_s}$, as shown in Fig. 5(h), indicates some scaling. $(\ln \tau)_{av}$ vs $\ln N$ is approximately in a straight line except the point of $N=100$. Simply linear fitting yields 
\begin{equation}
(\ln \tau)_{av}=-2.7(1)\ln N+8.0(5).
\end{equation}
Similarly,  $\delta (\ln \tau)$ vs $\ln N$ is almost in a straight line except the point of $N=100$. Simply linear fitting yields 
\begin{equation}
\delta (\ln \tau)=1.09(6)\ln N-3.5(3).
\end{equation}
The above two equations tell us  that $\tau$ distributes in larger scales for larger lattices although its average becomes smaller. 

These results tell us that the interfacial specific heat peak becomes higher and higher, narrower  and narrower, as the system size increases. It is expected that it becomes a $\delta$-function in the limit of $N\rightarrow \infty$. This just is the most important characteristic of the first order.

\begin{figure}
\includegraphics[width=0.5\textwidth]{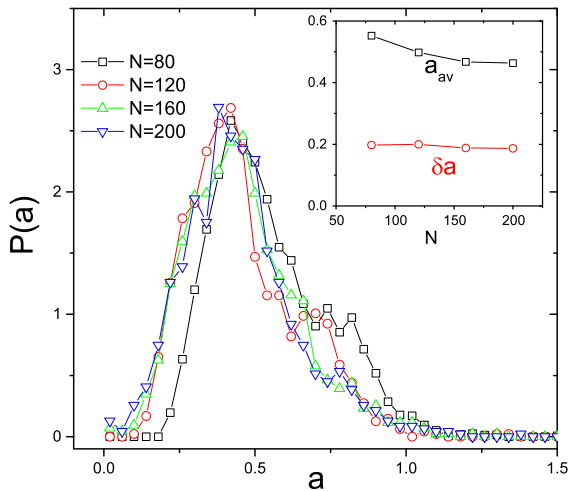}
\caption{ The distributions of $a=c_m \tau$, which characterizes the jump of the interfacial internal energy below and above the transition temperature, for $N=80,120,160,200$. The inset (h) shows the average of $a$ and it deviation $\delta a$ vs the system size}

\end{figure}

For the $i$th sample, the interfacial specific heat peak has a maximum $c_{mi}$ and a width $\tau_i$. We define their product $c_m \tau$ as
\begin{equation}
a_i=c_{mi} \tau_i
\end{equation}
which equals approximately the integration over the interfacial specific heat peak (divided by $\pi$). See Eq. (\ref{eq:gauss}). It equals the jump of the interfacial internal energy (divided by $\pi$) below and above the transition temperature approximately. See Fig 2(a), the jump of the interfacial internal energy at the transition point is different from sample to sample. We calculate the distributions $P(a)$ of $a$, which are shown in Fig. 8. 

As we can see, the distribution functions are in the similar shape. The averages $a_{av}$ seem to converge to a limit as the lattice size increases and the deviation $\delta a$ changes a bit as shown in the inset of Fig. 8.

The deviation of the transition temperature does not converge to zero as the lattice size increases shown in the inset of Fig. 5.  If this feature remains as the system size increase to infinity, the wetting transition temperature is sample dependent. Of course  we can not guarantee the validity of the extrapolating our results on the finite size lattices to the infinite size. However we will show a lot of evidence pointing to that the wetting transition temperature should be sample dependent in the thermodynamic limit.

\section{PHYSICAL PICTURE OF THE WETTING TRANSITION: COMPETITION BETWEEN THE GROUPS OF ADJACENT LINE DEFECTS}

To understand the present wetting transition,  we retrospect the the previous results on the wetting transitions of the two-dimensional Ising model.

With $a_n=1$ for all $n$,  the system is known as the Abraham's model \cite{abraham}, which is solved exactly by Abraham . It undergoes a continuous wetting transition at a temperature $T_w$ below the critical temperature $T_C$ of the 2D Ising model. For $T_w<T<T_C$, the interface is infinitely far from the left wall and the  interfacial free energy is obtained by Onsager \cite{onsager} 
\begin{equation}
 f_{O}=2k_BT (K-K^*)
\label{eq:onsager}
\end{equation}
where $K=J/k_B T$  and $\exp(-2K^*)=\tanh 2K$. For $T<T_w$, the interface is pinned at the the left wall and the interfacial free energy is obtained by Abraham \cite{abraham} 
\begin{equation}
 f_{A}=-k_B T\ln (A-\sqrt{A^2-1})
\label{eq:abraham}
\end{equation}
with $A=\frac{1}{2}(B+1/B)+1-\frac{1}{2}(S+1/S)$,  $B=\tanh K^* \coth K$, and $ S=e^{2K} (\cosh 2K-\cosh 2a_0 K)/\sinh 2K$. At the wetting transition point, it has
\begin{equation}
f_O(T_w)=f_A(T_w).
\end{equation}
At $T=T_w$ it has $ e^{2K}[\cosh 2K-\cosh 2a_0 K]=\sinh 2K$.

Forgacs {\sl et al.} found that the wetting transition is first-order if one adds a line defect in the bulk \cite{privman}, say the $N_1$th column bonds being $a_{N_1}J$, where $a_0<a_{N_1}<1$ and $1\ll N_1<N$. In the wetting phase, the interface is pinned at the line defect. The interfacial free energy is  given by
\begin{equation}
 f_{FSP}=-k_B T\ln |-x-\sqrt{x^2-1}|
\label{eq:x}
\end{equation}
where $x=(c c^*-c_2\sqrt{s_2^2 s^4+1})/(s^2_2 s^2-1)$ with $c=\cosh 2K$, $s=\sinh 2K$, $c^*=\cosh 2K^*$, $c_2=\cosh 2(K_2^*-K^*)$, $s_2=\sinh 2(K_2^*-K^*)$ and $K_2=a_{N_1}K$, $\exp(-2K_2^*)=\tanh 2K_2$.  Simply speaking, this first order wetting transition is a competition of two interface situations, one for the interface pinned at the left wall and one for the interface pinned at the line defect. For $T_1<T<T_C$, where $T_1$ is the first order wetting transition temperature \cite{privman}, it has $f_{FSP}<f_A$. Then, the interface is pinned at the defect line and the interfacial free energy is given by $f_{FSP}$.  On the contrary, for $T<T_1$, it has $f_{FSP}>f_{A}$.   Then the interface is pinned at the left wall and the interfacial free energy is given by $f_{A}$. The first-order transition temperature $T_1$ is given by
\begin{equation}
f_A(T_1)=f_{FSP}(T_1).
\end{equation}

In a sense the present model is an extension of Forgacs {\sl et al.}'s idea, i.e. adding many line defects into the system randomly. 
To understand the complicated cases of random bond, we consider the following imhomogeneous models with size $N=120$ 
\begin{eqnarray}
S2:~~~~a_n & =& 0.9~~~ for ~~~n=60,61; \nonumber \\
S4:~~~~a_n & =& 0.9~~~ for ~~~n=59,60,61,62. \nonumber \\
D24:~~~~a_n & = & 0.9~~~ for ~~~n=40,41,79,80,81,82; \nonumber \\
D42:~~~~a_n & = & 0.9~~~ for ~~~n=39,40,41,42,80,81; \nonumber 
\end{eqnarray}
and it still has $H_1=a_0=0.4$ and  $a_n=1.0$ for other $n$. In the case $S2$, there is a single group of two adjacent line defects located about $n=60$. In the case $S4$, there is a single group of four adjacent line defects located about $n=60$. In the case  $D24$, both the two groups of two adjacent line defects and four adjacent line defects are present,one group is located about $n=40$ and another one about $n=80$. In the case $D42$, the two groups swap their positions. We want to know how the two groups of adjacent line defects determine the wetting transition in the cases $D24$ and $D42$. The two cases $S2$ and $S4$ with a single group are studied to be references.

\begin{figure}
\includegraphics[width=0.5\textwidth]{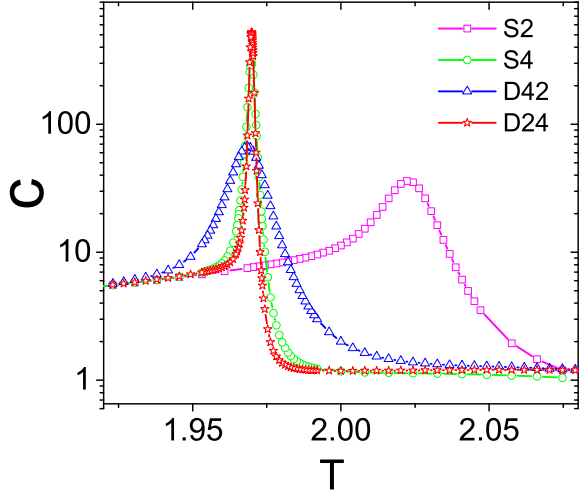}
\caption{ The interfacial specific heat  VS the temperature for the cases: S2,S4,D24 and D42.  }
\end{figure}

We calculate the interfacial specific heat, magnetization profile and the interface position for the four cases. The results are shown in Fig. 9, 10 and 11. We conclude that  the group of four adjacent line defects dominates the transition in the $D42$ and $D24$ cases. 

The first clue is that it has the same wetting transition temperature $T_w$ in cases $D24$, $D42$ and $S4$. As shown in Fig. 9(a), the wetting transition takes place at $T_w\approx 2.022$ in the $S2$ case, and at $T_w\approx 1.970$ in the $S4$ case. In both cases $D24$ and $42$, it has $T_w\approx 1.970$, which is the same as that in the case $S4$. 

The second clue is the interfacial free energy. As shown in Fig. 10, The interfacial free energies almost coincide in cases $D24$, $D42$ and $S4$. In Fig. 10, the solid lines in black and red are given by Eq. (\ref{eq:onsager}) and (\ref{eq:abraham}) respectively. The solid line in green (and blue) is the interfacial free energy in the $S2$ (and $S4$) case with $a_0=1$, which is related to model B in reference \cite{privman} and the interface is always pinned at the adjacent line defects. The interfacial free energy for $S2$ with $a_0=0.4$ is divided into two parts by the wetting transition point (marked by the blue arrow): it coincides with the solid line in red for $T<T_w$ and with the green line for $T>T_w$. The interfacial free energies in case $D24$ and $D42$  almost coincide with that in $S4$. They are divided into two parts by the wetting transition point marked by the red arrow: they coincide with the solid line in red for $T<T_w$  and with the solid line in blue for $T>T_w$.

\begin{figure}
\includegraphics[width=0.5\textwidth]{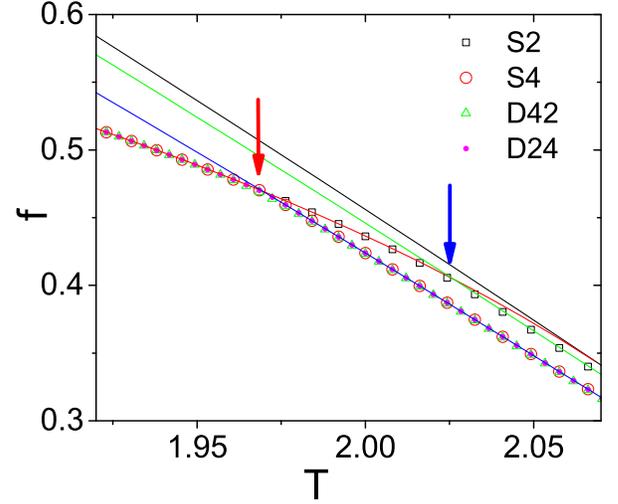}
\caption{ The interfacial free energy  VS the temperature for the cases: S2,S4,D24 and D42. See the text for the solid lines. }
\end{figure}

The third clue is that at the wet phase the interface is localized at the group of four adjacent line defects in both cases $D24$ and $D42$. This is a direct evidence that the group of four adjacent line defects dominates the wetting transition in the cases $D24$ and $D42$. At the temperature $T=1.961$, the system is in the nonwet phase for both $D24$ and $D42$ and the interface is pinned at the left wall as shown in Fig. 11(a). At the temperature $T=2.083$, the system is in the wet phase and the interface is pinned at the group of four adjacent line defects. For  $D24$ and $D42$, the interface is pinned at about $n=80$ and $n=40$ respectively, where the group of four adjacent line defects is located. Fig. 11(b) shows that above the wetting transition temperature $T_w=1.970$the interface position is pinned at the group of four adjacent line defects: $x_d\approx 80$ in the $D24$ case and $x_d \approx 40$ in the $D42$ case. Below the transition temperature the interface is pinned at the left wall, $x_d\approx 0$ in both cases $D42$ and $D24$.

\begin{figure}
\includegraphics[width=0.5\textwidth]{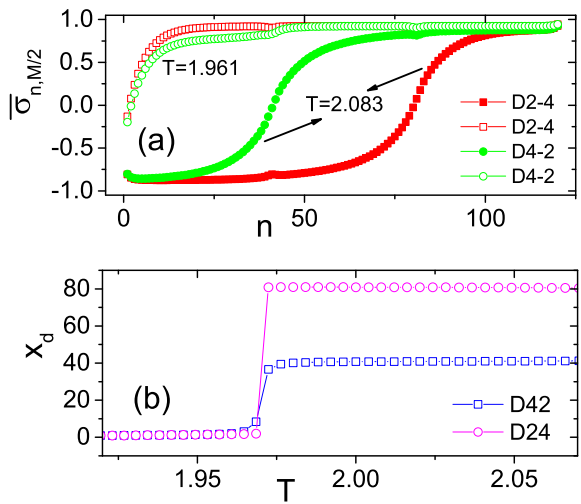}
\caption{ (a) The magnetization profile for the cases $D24$ and $D42$ at two temperatures. (b) The interface position v.s. the temperature for the cases: $D24$ and $D42$.  }
\end{figure}

Enlighten by  these imhomogeneous models, we conclude that the phase transition is the competition among the interface locations. See the Fig. 10, the interfacial free energy of the group of four  adjacent line defects in $S4$ case (in solid blue line) is lower than that of the two adjacent line defects in the $S2$ case (in solid green line) above the transition temperature. In both $D24$ and $D42$ cases, the interfaces  are pinned at the group of four adjacent line defects. Below the transition temperature the interfacial free energy at the left wall (in solid red line) is lower, then the interfaces jump from the group of four line defects to the left wall.

\section{Two semi-random models}

Extending the above discussion, one can conjecture that the wetting transition in the random bond model Eq. (1) should be the competition among interface locations. From the renormalization group theory, the adjacent line defects can be dealt as a single effective line defect. The more the adjacent line defects are, the weaker the bond of the effective line defect is. Therefore in the wet phase, the interface should be pinned at the group with the most adjacent line defects far from the left wall. The situations are so for the eight samples shown in Fig. 2.

For an infinite system, one can find a group with an arbitrarily large number of adjacent line defects. The lower bound of the interfacial free energy is given as the interface is pinned at the group with infinite line defects. It can be obtained from the Onsager's exact result \cite{onsager} and is given by $ f=2k_BT (0.9K-K^*)$, where $0.9K$ is the weak bond in the present model. As the system size goes to infinity, the interfacial free energy for the interface pinned at the group with large number of adjacent line defects should converge to this lower bound.

At the nonwet phase, the interface is pinned at the left wall.  See Fig. 4(a) and 11(a), the absolute value of magnetization is depressed notably only near the left wall. From Eq. (\ref{eq:interfacial}) we know that the interfacial free energy is defined as the free energy with boundary condition $+-$ subtracted by that with boundary condition $++$. Under the boundary condition $++$, the magnetization is close to $1$ because the system is deeply ordered. The interfacial free energy is definitely related to the difference of the magnetization between boundary condition $+-$ and $++$.  Just these differences induce the interfacial free energy.  The magnetization differences decay rapidly as the distance from the left wall increases. As we can see in Fig. 4(a) and 11(a), the absolute value of magnetization is about $0$ near the left wall, but approaches to $1$ as $n>20$.  The interfacial free energy is not only related  to the magnetization but also to the configuration of bonds. Obviously the bonds far from the the left wall, say $n>20$ in Fig. 11(a), are not closely related to the interfacial free energy. Only the bonds close to the left wall are related to the interfacial free energy since the interface is pinned at the left wall. In other words, the interfacial free energy for the interface pinned at the left wall should be related only to finite number column bonds near the left wall. Therefore as the system size goes to infinity, i.e. $N\rightarrow \infty$, the interfacial free energy for the interface pinned at the left wall should not converge to certain limit, because the interfacial free energy is only related to the configuration of the disorder near the left wall and not related to the disorder at other regions. Hence the spreading of the interfacial free energy for the interface pinned at the left wall should keep the same as $N\rightarrow \infty$.

\begin{figure}
\includegraphics[width=0.5\textwidth]{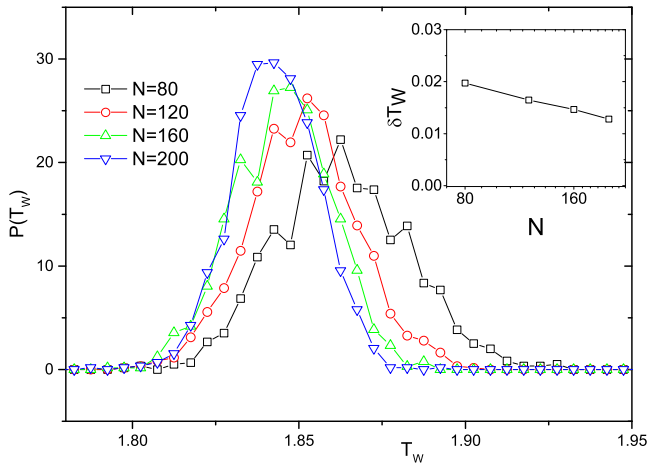}
\caption{ The distributing of wetting transition temperatures in the SR1 case with $N=80,120,160,200$. The inset shows the deviation of wetting transition temperatures.}
\end{figure}

The intersection between the interfacial free energies at the left wall and at the group with the most adjacent line defects determines the phase transition temperature. As discussed above the spreading of the interfacial free energy at the left wall will not  decreases as the size of the system goes to infinity, so the distribution width $\delta T_w$ will not converge to be zero as the usual phase  transition in the disordered systems.

To test this argument, we design two semi-random lattices. We call them SR1 and SR2. In SR1 case, we set  that $a_n=1.0$ for $n\le 20$ and $a_n$ is random for $n> 20$. In other words, the bonds near the left wall are nonrandom for $n<20$. In  case SR2, we set that $a_n$ is random for $n \le 20 $, $a_n=0.9$ for $N/2< n \le N/2+20$, and $a_n=1.0$ for other $n$. In this case only the bonds near the left wall have randomness. We solve the model on these two semi-random lattices for $N=80,120,160,200$ and study the distribution of $T_w$ over more than $1000$ samples. The distributions of $T_w$ for SR1 and SR2 case are shown in Fig. 12 and 13 respectively. In the SR1 case, the number of samples are $1197,1221,1292,1174$ for $N=80,120,160,200$ respectively. In the SR2 case, the number of samples are $1598,1193,1141,1231$ for $N=80,120,160,200$ respectively.

\begin{figure}
\includegraphics[width=0.5\textwidth]{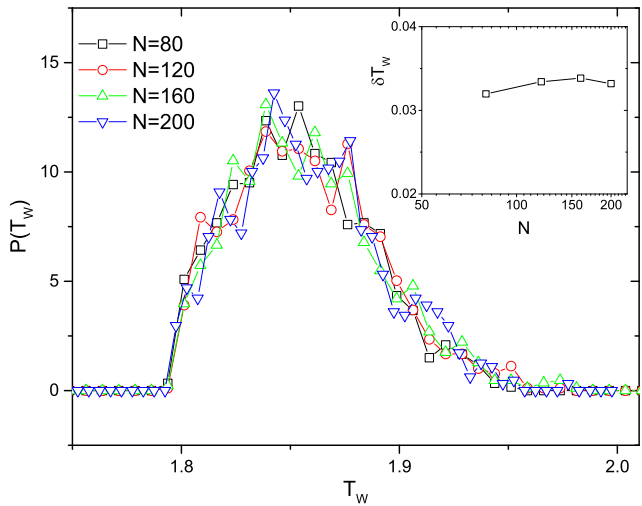}
\caption{ The distributing of wetting transition temperatures in the SR2 case with $N=80,120,160,200$. The inset shows the deviation of wetting transition temperatures.}
\end{figure}

In the SR1 case, the interfacial free energy at the left wall is fixed since the bonds near the left wall are nonrandom and fixed. As the system size increases, the maximum of the number of the adjacent line defects increases. Then the interfacial free energy for the interface pinned at the group with most adjacent line defects will converge to the  limit of infinite line defects. Therefore the distribution width $\delta T_w$  will decrease as the system size $N$ increases. In Fig. 12, the distribution width $\delta T_w$ is $1.973\times 10^{-2}$, $1.648\times 10^{-2}$, $1.467\times 10^{-2}$,$1.282\times 10^{-2}$ for $N=80,120,160,200$ respectively.   The deviation of $T_w$ indeed shows the trend of converging to zero as the lattice size increases. 

On the contrary, in the SR2 case the interfacial free energy at the left wall will not converge as the system size increase, so the distribution width $\delta T_w(N)$ will not decrease. The numerical results for SR2 is shown in Fig. 13, in which the distribution width $\delta T_w$ is $3.197\times 10^{-2}$, $3.342 \times 10^{-2}$, $3.384 \times 10^{-2}$, $3.320\times 10^{-2}$ for $N=80,120,160,200$ respectively. It indeed does not decrease as the lattice size increases. The numerical results on the two semi-random models  are consistent to our expectation.

\section{DISCUSSION}
 
It is very unusual that the wetting transition temperature is sample dependent. For the usual phase transition in the disordered systems, the phase transition temperatures converge to a limit as the system size goes to infinity. For the usual phase transition, the free energy is related to the whole system. As first argued by Brout \cite{brout}, we may divide the system into n large subsystems (much larger than the correlation length). If we assume that the coupling between neighboring subsystems is negligible, then the value of any density of an extensive quantity over the whole sample is equal to the average of the (independent) values of this quantity over the subsystems. The pseudo-phase-transition temperature fluctuates from sample to sample due to the finite-size effects. However as the system size goes to infinity, the pseudo-phase-transition temperatures should converge to a limit $T_C(\infty)$ \cite{aharony}. In the present wetting transition  the interfacial free energy is only related to the left wall and the group of the most adjacent line defects. Obviously, the present model can not  be divided into two similar subsystems in the horizontal direction. If the left and right wall are separated, there is no wetting transition.

The groups of adjacent line defects are the so-called rare regions \cite{vojta}. Near the critical point of the McCoy-Wu model, the rare regions dominate the phase transition \cite{vojta,fisher}. In this wetting transition, the situation is more extreme. Only the largest rare region matters. 

In fact, the wetting transition in the random bond systems has been also studied extensively \cite{kadar,lipowsky,kadar1,lipowsky1,huang,wuttke}. However in these previous studies, the random bonds are not correlated. In the McCoy-Wu Ising model the random bonds are perfectly correlated in one direction \cite{mccoy}. This should be the main reason which makes this transition  so different.

Because the McCoy-Wu model is equivalent to the one-dimensional random transverse field quantum  Ising model \cite{fisher}, it is expected that a similar wetting transition exists in the one-dimensional random transverse field quantum  Ising model. The quantum Ising chains with boundary fields has been studied by Campastrini 
et. al \cite{campastrini}. There is a magnet-to-kink transition similar to the critical wetting transition in the Abraham model \cite{abraham}. The random quantum Ising 
chain with boundary fields is the quantum version of the wetting transition in the McCoy-Wu model.

It is pleasure to thank professor Kurt Binder for useful discussions. The author also thanks professor Wenan Guo and the SGI in Department of Physics in Beijing Normal University for the supply of computing time.


\begin{thebibliography}{}
\bibitem{vojta}Thomas Vojta, J. Phys. A: Math. Gen. {\textbf 39}, R143 (2006).  

\bibitem{diluteising}Ballesteros H G, Fernandez L A, Martin-Mayor V and Sudupe A M  Phys. Rev. B {\textbf 58}, 2740 (1998) 

\bibitem{landau}Ferrenberg A M and Landau D P, Phys. Rev. B {\textbf 44}, 5081 (1991).

\bibitem{mccoy}McCoy B M and Wu T T  Phys. Rev. Lett. {\textbf 21,} 549 (1968) 

\bibitem{mccoy1}B. M. McCoy, Phys. Rev. Lett.23, 383 (1969).

\bibitem{fisher}D. S. Fisher,Phys. Rev. Lett.69, 534 (1992).  D. S. Fisher,Phys. Rev. B51, 6411 (1995).  

\bibitem{aizenman}M. Aizenman, J. Wehr, Phys. Rev. Lett., {\textbf 62}, 2503 (1989). 

\bibitem{domany}S. Wiseman and E. Domany, Phys. Rev. E, {\textbf 52}, 3469 (1995).

\bibitem{bellafard}A. Bellafarda, S. Chakravartya, M. Troyerb, H. G. Katzgraberc, Ann. Phys., {\textbf 357}, 66 (2015).

\bibitem{chakravarty}A. Bellafard and S. Chakravarty, Phys. Rev. B, {\textbf 94}, 094408 (2016). 

\bibitem{abraham}D. B. Abraham, Phys. Rev. Letts., {\textbf 44}, 1165  (1980).

\bibitem{privman} G. Forgacs, N. M. $\breve{S}$vraki$\acute{c}$, and V. Privman, Phys. Rev. B {\textbf 38}, 8996 (1988).

\bibitem{loh}Y.L.Loh and E.W.Carlson, Phys.Rev.Lett.{\textbf 97}, 227205 (2006).  

\bibitem{wu1}X. Wu, R. Zheng, N. Izmailian and W. Guo, J. Stat. Phys. {\textbf 157}, 1284 (2014). 

\bibitem{wu} X. Wu, J. Stat. Phys. {\textbf 155}, 106 (2014). 

\bibitem{binder}E. V. Albano, K. Binder, D. W. Heermann, and W. Paul, J. Stat. Phys., {\textbf 61}, 161 (1990). 

\bibitem{binder1} K. Binder, Rep. Prog. Phys., {\textbf 50}, 783 (1987).

\bibitem{onsager} L. Onsager, Phys. Rev. {\textbf 65}, 117 (1944). 

\bibitem{brout} R. Brout, Phys. Rev. {\textbf 115}, 824 (1959).

\bibitem{aharony}A. Aharony and A. B. Harris, Phys. Rev. Lett., {\textbf 77}, 3700 (1996).

\bibitem{campastrini}M. Campostrini, A. Pelissetto  and E. Vicari, J. Stat. Phys., P11015 (2015).

\bibitem{kadar} M. Kardar, Phys. Rev. Lett., {\textbf 55}, 2235 (1985). 

\bibitem{lipowsky}R. Lipowsky and M. E. Fisher, Phys. Rev. Lett. 56, 472 (1986). R. Lipowsky and M. E. Fisher, Phys. Rev. B36, 2126 (1987). 

\bibitem{kadar1} M. Kardar, Nucl. Phys. B 290, 582 (1987).

\bibitem{lipowsky1}R. Lipowsky, Phys. Scr., T 29, 259 (1989). 

\bibitem{huang} M. Huang, M. E. Fisher, and R. Lipowsky, Phys. Rev. B 39, 2632 (1989). 

\bibitem{wuttke} J. Wuttke and R. Lipowsky, Phys. Rev. B 44, 13042 (1991). 



\end{thebibliography}
\end{document}